\documentclass[preprint,aps,superscriptaddress,showkeys,showpacs,pre]%
{revtex4}
\usepackage{amsmath,amssymb}
\usepackage{epsfig}
\usepackage{graphics}
\usepackage{graphicx}
\usepackage{float}
\usepackage{dcolumn}
\begin{document}

\title{STATISTICAL MEASURES AND DIFFUSION DYNAMICS IN A MODIFIED CHUA'S  CIRCUIT EQUATION WITH MULTI-SCROLL ATTRACTORS }

\author{ G.~SAKTHIVEL, S.~RAJASEKAR$^\dagger$\\
School  of Physics, Bharathidasan University, \\ Tiruchirapalli 620 024, Tamilnadu, India \\ $^\dagger$rajasekar@cnld.bdu.ac.in }

\author{\\ K.~THAMILMARAN \\
Centre for Nonlinear Dynamics, 
School  of Physics, \\Bharathidasan University,  Tiruchirapalli 620 024, \\ Tamilnadu, India \\ maran.cnld@gmail.com }

\author{\\ SYAMAL KUMAR DANA \\
Central Instrumentation, Indian Institute of Chemical Biology, \\
              Jadavpur, Kolkata 700 032, India \\
                     skdana@iicb.res.in \\}

\begin{abstract}
In this paper the focus is set on a modified Chua's circuit model equation with saw-tooth function in place of piece-wise linear function of Chua's circuit displaying multi-scroll chaotic attractors.  We study  the characteristic properties of first passage times ($t_\mathrm{FPT}$s) to $n$th scroll chaotic attractor, residence times ($t_\mathrm{RT}$s) on a scroll attractor and returned times ($t_\mathrm{RET}$s) to the middle-scroll attractor.  $t_\mathrm{FPT}$s exhibit a series of Gaussian-like distribution followed by a long tail continuous distribution.  $t_\mathrm{RT}$s and  $t_\mathrm{RET}$s show completely  discrete distribution.  Power-law variation of mean values of $t_\mathrm{FPT}$s, $t_\mathrm{RT}$s and $t_\mathrm{RET}$s with a control parameter is found.  On the other hand, mean values of $t_\mathrm{FPT}$s  and $t_\mathrm{RET}$s  have linear dependence with the number of the  scroll attractors for fixed values of the control parameter.  For the system with infinite scroll chaotic attractors  normal diffusive motion occurs.  In the normal diffusion process the mean square displacement grows linearly with time.


\end{abstract}
\maketitle

\section{Introduction}
Chua's circuit is found to show a rich variety of both local and global bifurcations and various types of chaotic  orbits [Madan, 1993; Tsuneda, 2005; Bilotta {\emph{et al.}}, 2007].  One of the  interesting chaotic orbits observed in the Chua's circuit is a double-scroll orbit.  Several modified Chua's circuit models have been proposed for generating multi-scroll chaotic attractors [Lu,  \& Chen, 2006].  For example, a three-segment piece-wise linear function $f$ characterizing  voltage-current characteristic of the Chua's diode in the Chua's circuit is replaced by a discontinuous function [Khibnik {\emph{et al.}}, 1993; Lamarque {\emph{et al.}}, 1999], sigmoid function [Mahla \& Badan  Palhares, 1993], multiple piece-wise linear segments [Suykens \& Vandewalle, 1993; Aziz-Alaoui, 1999; Zhong  {\emph{et al.}}, 2002], sign function [Yalcin {\emph{et al.}}, 2001],  sine function [Tang {\emph{et al.}}, 2001], hyperbolic tangent function [Ozoguz {\emph{et al.}}, 2002], nonlinear term $x \vert x \vert$ [Tang {\emph{et al.}}, 2003], saturated function [Lu {\emph{et al.}}, 2004] and saw-tooth function [Yu {\emph{et al.}}, 2007a].  Experimental circuit design for generating one-dimensional [Yu {\emph{et al.}}, 2003], two-dimensional [Munoz-Pacheco \& Tlelo-Cuautle, 2009]   scroll  attractors, $(2m+1) \times (2m+1)$ couple of grid multi-scroll attractors [ Yu {\emph{et al.}}, 2006], hyperchaotic $n$-scroll  attractor [ Yu {\emph{et al.}}, 2007b] and  a systematic  methodology to design  circuits to achieve  desired swings, widths, slopes, breakpoints, equilibrium points and shapes of $n$-scroll chaotic attractors [Yu {\emph{et al.}}, 2005] have been proposed. Unidirectionally or  diffusively coupling [Kapitaniak \& Chua, 1994; Dana {\emph{et al.}}, 2008], cellular neural networks [Arena {\emph{et al.}}, 1996; Suykens \& Chua, 1997], a jerk circuit with Josephson junctions [Yalcin, 2007] and  thresholding approach [Lu {\emph{et al.}}, 2008] have also been used to generate multi-scroll attractors.  In a very recent work Campos-Canton et al. [Campos-Canton {\emph{et al}}., 2010] constructed a class of three-dimensional  dissipative unstable systems exhibiting multi-scroll chaotic attractors.  Practical applications of multi-scroll dynamics are found in broadband signal generators and pseudo-random number generator for communication engineering.

The basic ingredient of multi-scroll attractors in the modified Chua's circuits is the presence of multiple breakpoints in the function $f$.  Suppose the number of breakpoints in the function $f$ in a multi-scroll Chua's circuit model is very large in a practical consideration and theoretically infinite.  In such a case $f(x)$ is periodic over the entire range of the argument $x$.  The system can exhibit infinite number of scroll chaotic attractors with diffusion characteristics.  In this connection we note that certain nonlinear systems with periodic potential  exhibit diffusion dynamics [Blackburn \& Gronbech-Jensen, 1996; Popescu {\emph{et al}}., 1998; Harish {\emph{et al}}., 2002; Sakthivel \& Rajasekar, 2010].  When the mean square displacement, $\langle x^2 \rangle$, is proportional to $t^\mu$, $\mu > 0$ the underlying motion is called diffusion.  $\mu=1$ and $\mu \ne 1$ correspond to normal and anomalous diffusions.  Transition from normal to anomalous and vice-versa are found to occur in certain Hamiltonian systems and volume preserving maps [Zaslavsky, 2002].

Great deal of interest has been focused on the circuit design, identification of different forms of the piece-wise linear function $f(x)$ and shapes of various types of scroll attractors generated.  It is also important to analyse statistical dynamics of multi-scroll chaotic attractors and also use nonlinear  circuit models to explore statistical  properties associated with the dynamics exhibited by them.  For example,  in the multi-scroll chaotic systems, because a trajectory escapes from one scroll attractor to nearby scroll attractors, the first passage time (FPT), residence time (RT) and returned time (RET) of  scroll attractors are important.  The focus of the present work is to study the features of these quantities and diffusion dynamics in a modified Chua's circuit model equation with multi-scroll chaotic attractors.   The FPT ($t_\mathrm{FPT}$) is the time when a stochastic process $x(t)$ started at $t=0$ from a given initial value within a domain $\triangle$ of its state space crosses it first time.  For systems exhibiting multi-scroll chaotic attractors or escape dynamics from one part of an attractor to another part of the attractor or from one well of a potential to another well FPT and mean FPT are important and study of them has practical applications in many problems [Redner, 2001].  RT ($t_\mathrm{RT}$) of a scroll attractor is defined as the time duration spend by a trajectory on it before passing to another scroll attractor.  RET ($t_\mathrm{RET}$) of  a scroll attractor is the time taken by the system's trajectory to re-enter into it.  

\begin{figure}[b]
\begin{center}
\epsfig{figure=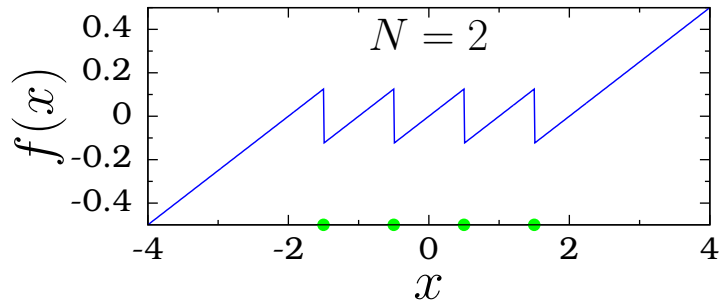, width=0.6\columnwidth}
\end{center}
\caption{The graph of  $f(x)$ given by Eq.~(\ref{eq1}d) with $N=2$, $\xi=0.25$ and $A=0.5$. The breakpoints of $f(x)$ are indicated by painted circles. }
 \label{fig1}
\end{figure}

We consider the modified Chua's circuit system [Yu {\emph{et al.}}, 2007]
\begin{subequations}
 \label{eq1}
\begin{eqnarray}
  \dot{x} & = & \alpha [ y - f(x) ] , \\
  \dot{y} & = & x - y + z , \\
  \dot{z} & = & - \beta y ,
\end{eqnarray}
where
\begin{eqnarray}
   f(x) & = & \xi x  - \xi A \sum_{j=0}^{N-1} {\bigg{\{}}
                {\mathrm{sgn}} {\big{[}}x+ (2 j+1) A {\big{]}}
                 +  {\mathrm{sgn}} {\big{[}}x- (2 j+1) A {\big{]}}
                     {\bigg{\}}}
\end{eqnarray}
with $\alpha, \, \beta, \, \xi, \, A>0$, $N \ge 1$ and 
\begin{eqnarray}
  {\mathrm{sgn}} [x] 
     & = & \left\{ \begin{array}{rl}
             1, &  \; \mathrm{if} \; x>0 \\ 
             0, &  \; \mathrm{if} \; x=0 \\ 
            -1, &  \; \mathrm{if} \; x<0 . \end{array}  \right.
\end{eqnarray}
\end{subequations}
$f(x)$ is a saw-tooth function with amplitude $2 A \xi$ and period $2A$.  Figure \ref{fig1} depicts the form of $f(x)$ for $N=2$, $\xi=0.25$ and $A=0.5$.  The breakpoints  $x^*$ of $f(x)$ are indicated by painted circles.  System (\ref{eq1}) is capable of generating $(2N+1)$-scroll attractors for a range of fixed values of the parameters $\alpha$ and $\beta$.  In the present work, on the system (\ref{eq1}) we consider the distributions  and mean values of FPTs (on the breakpoints $x^*$), residence times (on a scroll attractor) and return times (to the middle-scroll attractor) in the case of $(2N+1)$-scroll chaotic attractors.  We study the variation of these quantities with the parameter $\alpha$ for $N=1$ and with $N$ for few fixed values of $\alpha$.  
The probability distribution $P(t_{\mathrm{FPT}})$ exhibits a sequence of peaks followed by a long tail continuous distribution.  Power-law variation of  mean FPT with $\alpha$ and linear variation of it with $N$ for fixed values of $\alpha$ are found.  $t_{\mathrm{RT}}$s of scroll attractors are discrete.  The mean RT also shows power-law dependence on $\alpha$. For a fixed value of $\alpha$, the mean RT  on each attractors are all almost same and independent of $N$.  Further, $P(t_{\mathrm{RT}})$ has only a sequence of peaks.    On the other hand,  $P(t_{\mathrm{RET}})$  of middle-scroll attractor has a discrete distribution with number of Gaussian-like envelop. The number of such envelops increases with increase in the value of $N$.  The mean RET varies linearly with $N$ while it decays following a power-law relation with $\alpha$.  When $N=\infty$ (very large for a practical consideration)  the motion is diffusion along $x$-direction.  The mean square displacement of the $x$-component varies linearly with time implying normal diffusion and is further confirmed by kurtosis.
\section{Characteristics of FPT, RT and RET}
\begin{figure}[t]
\begin{center}
\epsfig{figure=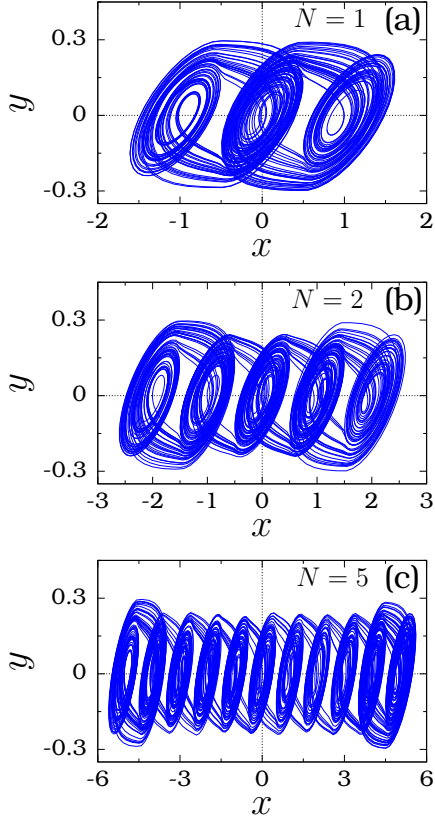, width=0.37\columnwidth}
\end{center}
\caption{Phase portrait of (a) three-scroll, (b) five-scroll and (c) eleven-scroll chaotic attractors  in the $x-y$ plane.  The values of the parameters in Eq.~(\ref{eq1}) are $\alpha=10$, $\beta=16$, $\xi=0.25$ and $A=0.5$. }
 \label{fig2}
\end{figure}

For our numerical study we fix the values of the parameters in Eq.~(\ref{eq1}) as $\xi=0.25$, $A=0.5$, $\beta=16$ and vary the parameter $\alpha$.   The $f(x)$ curve has discontinuities or breakpoints at
\begin{equation}
 \label{eq2}
   x^* = \pm \left( m - \frac{1}{2} \right), \quad m=1,2,\cdots, N.
\end{equation}
The equilibrium points of the system (\ref{eq1}) are given by 
\begin{equation}
 \label{eq3}
   (x_{\mathrm{e}}, y_{\mathrm{e}}, z_{\mathrm{e}} ) 
   = (0,0,0), \; ( \pm m,0,\mp m), \quad m=1,2,\cdots,N.
\end{equation}
The stability determining  eigenvalues of the equilibrium points  are the roots of the cubic equation
\begin{equation}
 \label{eq4}
   \lambda^3 + (1+ \alpha f') \lambda^2 
    + (\alpha f' + \beta - \alpha ) \lambda 
    + \alpha \beta f' =0, 
\end{equation}
where $f' = {\displaystyle{\frac{{\mathrm{d}}f}{{\mathrm{d}}x} {\Big{|}}_{x=x_{\mathrm{e}}}}} =0.25 $.  For $\alpha < \alpha_{\mathrm{c}}=7.74518$ all the equilibrium points are stable focus  with one eigenvalue being real negative and other two being complex conjugate with negative real part.  As $\alpha$ increases from a small value the magnitude of the negative eigenvalue increases while the magnitude of the real part of  the complex conjugate eigenvalues decreases.  At $\alpha=\alpha_{\mathrm{c}}$ the two complex conjugate eigenvalues  become pure imaginary.  For $\alpha > \alpha_{\mathrm{c}}$ the real part of the these two complex conjugate eigenvalues become  positive.  So, for $\alpha > \alpha_{\mathrm{c}}$ the equilibrium points are unstable.  In the numerical simulation we found sudden occurrence of chaotic motion (crisis) at  $\alpha = \alpha_{\mathrm{c}}$.  Figure \ref{fig2} shows the phase portrait of the scroll chaotic attractors in $x-y$ plane for $\alpha=10$ and for three values of $N$.   

%

\begin{figure}[t]
\begin{center}
\epsfig{figure=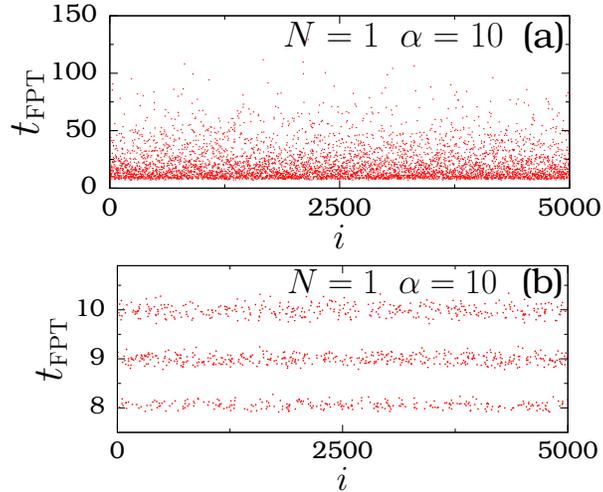, width=0.5\columnwidth}
\end{center}
\caption{(a) Numerically computed FPTs on the barrier $x^*=0.5$ of the system (\ref{eq1}) for $N=1$ and $\alpha=10$.  (b) Magnification of part of (a) showing finer details.  }
 \label{fig3}
\end{figure}

\begin{figure}[t]
\begin{center}
\epsfig{figure=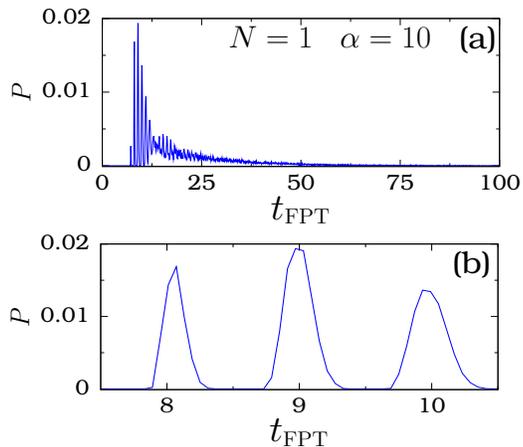, width=0.44\columnwidth}
\end{center}
\caption{(a) Plot of probability distribution of $t_{\mathrm{FPT}}$ on the barrier $x^*=0.5$ with $N=1$ and $\alpha=10$.  (b) Magnification of part of $P$ in (a) corresponding to the Fig.~\ref{fig3}(b).}
 \label{fig4}
\end{figure}

Now, we present the features of FPT, mean FPT, RT, mean RT, RET and mean RET.  First we consider the case $N=1$.    We calculate the FPT on the barrier (breakpoint) $x^*$ for a set of random initial conditions chosen around the origin for $\alpha>\alpha_{\mathrm{c}}$.  Figure \ref{fig3}(a) shows the numerically computed $t_{\mathrm{FPT}}$ on $x^*=0.5$ for $5000$ initial conditions with $\alpha=10$.  $t_{\mathrm{FPT}}$ depends on initial condition.  $t_{\mathrm{FPT}}$s are distributed over a range.  Figure \ref{fig3}(b) shows the magnification of part of Fig.~\ref{fig3}(a).  The probability distribution $P(t_{\mathrm{FPT}})$ is calculated using $10^5$ $t_{\mathrm{FPT}}$s and is depicted in Fig.~\ref{fig4}.  $P$ is not a monotonically increasing or decreasing function of FPT.  It has multiple peaks and a very long tail.  The height of the peaks first increases with increase in the value of FPT, reaches a maximum value and then slowly  decays  for larger values of FPT.  $P$ is continuous for larger values of FPT.  For  smaller values of $t_{\mathrm{FPT}}$ the distribution $P$ shows a finite number of Gaussian-like profile.  This is clearly seen in Figs.~\ref{fig3}(b) and \ref{fig4}(b).  We note that the peaks of the distribution occur at regular interval of $t_{\mathrm{FPT}}$.  Between first few bands there exists a range of time interval within which almost no $t_{\mathrm{FPT}}$ occurs.  The above character of $t_{\mathrm{FPT}}$ is found for the barrier $x^*=-0.5$ and also for other values of $\alpha$.

\begin{figure}[!h]
\begin{center}
\epsfig{figure=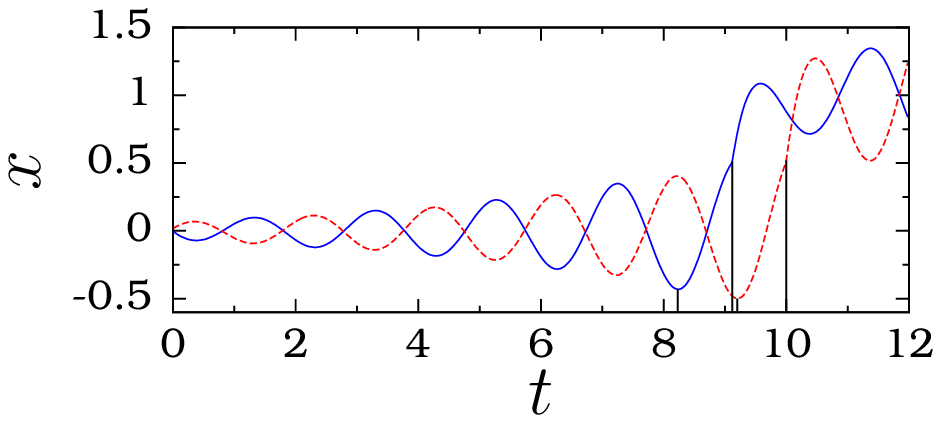, width=0.48\columnwidth}
\end{center}
\caption{$x(t) $ versus $t$ for two trajectories showing the time of exit from the middle-scroll attractor.  The breakpoint is $x^*=0.5$.  }
 \label{fig5}
\end{figure}

In Fig.~\ref{fig2} we notice that trajectories leave a scroll attractor through a specific confined exit regions.  Consider the middle-scroll confined between the breakpoints $x^*=-0.5$ and $0.5$.  Through one exit region trajectories enter into left-scroll while through the other they enter into right-scroll.   A trajectory starting from the vicinity of an exit takes a minimum time to arrive at the neighborhood of the next exit.  This minimum time is found  to be nearly unity.  Figure \ref{fig5} shows two trajectories started near the origin and leaving the middle-scroll after some time.  In this figure for each trajectory there are two vertical lines: the right-line denotes the time at which the trajectory  leaves the middle-scroll and the left-line marks the latest earlier  time at which $x(t)$ came near an exit region  before leaving the scroll.  The time difference between the two vertical lines is roughly unity.

\begin{figure}[b]
\begin{center}
\epsfig{figure=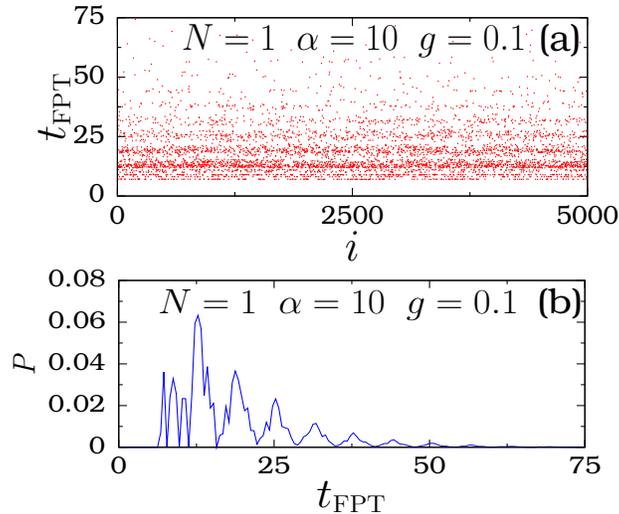, width=0.5\columnwidth}
\end{center}
\caption{$5000$ FPTs obtained for the distance $x^*=0.5$ and the probability distribution of FPT for the system (\ref{eq1}) driven by the periodic force $g \cos t$ with $g=0.1$. }
 \label{fig6}
\end{figure}

It is interesting to point out the effect of addition of external periodic force on FPT. Suppose the system (\ref{eq1}) is  driven by the periodic force $g \cos t$.  We add this external force to Eq.~(\ref{eq1}a).  Figure \ref{fig6} shows $t_{\mathrm{FPT}}$ and $P(t_{\mathrm{FPT}})$ for two values of $g$.  
The continuous and nonGaussian distribution  observed in Fig.~\ref{fig4}(a) for $t_{\mathrm{FPT}}>12$ disappear and a series of Gaussian-like distribution occurs even for small values of $g$.  This is evident in Fig.~\ref{fig6}(b) where $g=0.1$.   The number of peaks decreases with increase in the value of $g$ and further  $P$ becomes more and more  narrow and   the peaks occur at regular interval of time.  The time intervals between successive peaks of $P$ are almost same and $\approx 2\pi$, the period of the external force $g \cos t$.   The periodic nature of  $P(t_{\mathrm{FPT}})$ can be easily understood.  The external force periodically oscillates the barrier heights of the function $f(x)$.  The chance for crossing the critical value $x^*$ is large when the barrier height at $x^*$ becomes a minimum which happens once during every period of the external periodic force.  This is the reason for the periodic occurrence of peaks.  If the  periodic force is replaced by a Gaussian white noise  then  the FPTs become random, the band-like structure would disappear and $P$ becomes continuous.  In a purely noise induced process the FPT distribution goes through a single maximum and then decays following either exponential or power-law relation.   One can expect multi-peaks in the presence of both periodic force and weak noise.  Suppression of chaotic motion by the addition of weak periodic force and delay feedback is realized in several chaotic systems [Pyragas, 1992; Braiman \& Goldhirsch, 1991].  In the system (\ref{eq1}) the effect of added periodic force is numerically studied for $\alpha \in [0,10]$ and $g \in [0,0.5]$.  For $\alpha > \alpha_{{\mathrm{c}}}$ suppression of chaos is not observed.  For $\alpha < \alpha_{{\mathrm{c}}}$ period-$T(=2 \pi)$ orbit is found for the above chosen range of values of $g$.  For each fixed value of $\alpha < \alpha_{\mathrm{c}}$, for $g$ value  less than a critical value there are three coexisting stable periodic orbits.  They occur about the equilibrium points of the system.  As $g$ increases from a small value the size of the orbit increases and above a certain critical value the system has a single period-$T$ orbit enclosing all the three equilibrium points.  Any route to chaotic motion  is not found  for the parametric choices considered here. 

\begin{figure}[t]
\begin{center}
\epsfig{figure=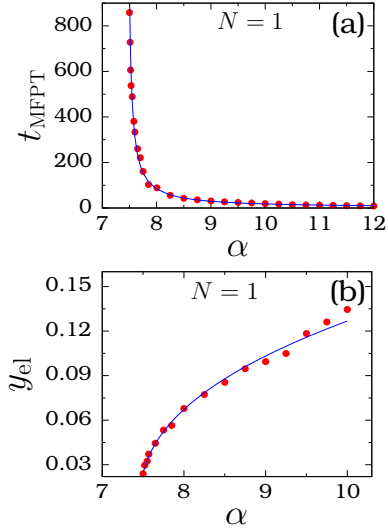, width=0.32\columnwidth}
\end{center}
\caption{(a) Variation of mean FPT with the parameter $\alpha$ for the target location $x^*=0.5$.  Painted circles are numerical data while the continuous line is the best power-law fit. (b) Power-law dependence of $y_{{\mathrm{el}}}$ with $\alpha-\alpha_{{\mathrm{c}}}$. }
 \label{fig7}
\end{figure}

The mean FPT ($t_{\mathrm{MFPT}}$), the mean time the trajectory initially in the neighbourhood of the origin takes to cross the target location $x^*$ first time, is calculated for a range of values of $\alpha > \alpha_{\mathrm{c}}$.  It is obtained averaging over $10^5$ FPTs.  Numerically computed $t_{\mathrm{MFPT}}$ versus $\alpha$ is plotted in Fig.~\ref{fig7}(a) with $x^*=0.5$.  It decreases rapidly with $\alpha$ following the power-law relation: $t_{\mathrm{MFPT}} = 45.688 (\alpha - \alpha_{\mathrm{c}} )^{-1.003}$, i.e., $t_{\mathrm{MFPT}} \propto 1/(\alpha - \alpha_{\mathrm{c}})$.  In Fig.~\ref{fig2} we can clearly see that trajectory moving on a scroll attractor leaves it  when enters into a certain interval of $y$ called turnstile.  We call  the length of this interval of $y$ as exit length and denote it as $y_{{\mathrm{el}}}$.  This quantity for the middle-scroll attractor is numerically calculated for a range of values of $\alpha$ and the result is shown in Fig.~\ref{fig7}(b).  $y_{{\mathrm{el}}}$ is found to increase with $\alpha-\alpha_{{\mathrm{c}}}$ obeying the power-law relation $y_{{\mathrm{el}}}=0.08618(\alpha-\alpha_{{\mathrm{c}}})^{0.4127}$.  The power-law decay of $t_{\mathrm{MFPT}}$ with the parameter $\alpha$ in Fig.~\ref{fig7}(a) can be due to the power-law increase in  $y_{{\mathrm{el}}}$ with $\alpha$.

The characteristics of FPT on $x^*=0.5$  for $N=1$ are found for other values of $N$ and $x^*$.    In all the cases band-like patterns with gap between the bands for smaller values of FPT and continuous  distribution for larger values are observed.   The range of FPT is found to increase with increase in the value of $N$.  This is because the region accessible to the trajectories in the region $x<0$ increases with increase in the value of  $N$.

\begin{figure}[t]
\begin{center}
\epsfig{figure=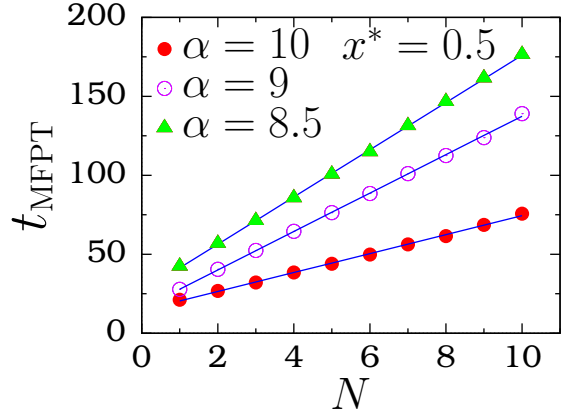, width=0.45\columnwidth}
\end{center}
\caption{MFPT versus $N$ for three fixed values of $\alpha$.  Symbols are numerical data and continuous lines are the best straight-line fits. }
 \label{fig8}
\end{figure}

We analyse the dependence of $t_{\mathrm{MFPT}}$ (on the target $x^*=0.5$) with $N$, the number of the saw-tooth function.  In Fig.~\ref{fig8} we plotted $t_{\mathrm{MFPT}}$ versus $N$ for three fixed values of $\alpha$.  Linear variation of $t_{\mathrm{MFPT}}$ with $N$ is observed for each fixed value of $\alpha$.  We obtained
%
\begin{equation}
 \label{eq5}
   t_{\mathrm{MFPT}} (x^*=0.5) = \begin{cases}
        14.955 N + 26.46, \;\; \alpha=8.5 \\
        12.167 N + 15.753, \;\; \alpha=9 \\
        5.996 N + 14.459, \;\; \alpha=10.
        \end{cases}
\end{equation}
In Fig.~\ref{fig8} we notice that for a fixed value of $N$ the  $t_{\mathrm{MFPT}}$ decreases with increase in $\alpha$.  The variation of $t_{\mathrm{MFPT}}$ with $\alpha$ follows  power-law relation while it with $N$ follows linear relation.  The rate of variation of MFPT with $N$ decreases with increase in $\alpha$.  

\begin{figure}[t]
\begin{center}
\epsfig{figure=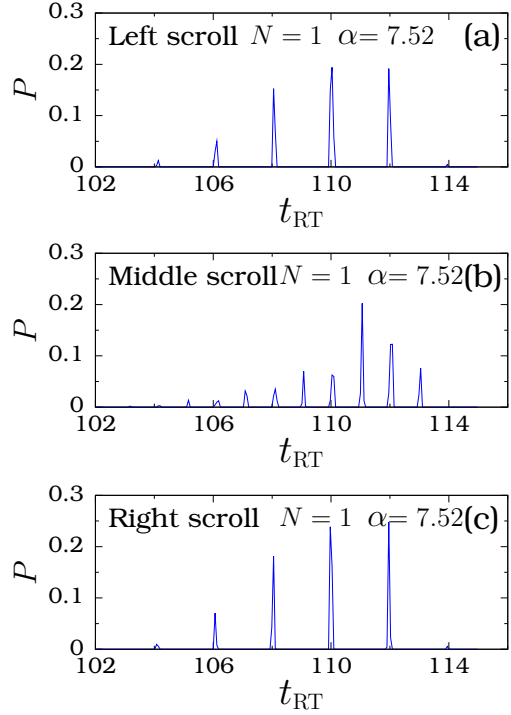, width=0.43\columnwidth}
\end{center}
\caption{Plots of probability distribution of $t_{\mathrm{RT}}$ of the three scroll attractors of the system (\ref{eq1}) for $N=1$ and $\alpha=7.52$.    }
 \label{fig9}
\end{figure}

Next, we wish to present the results on residence time statistics.  We denote $t_{\mathrm{RT}}$ as the time the trajectory resides on a scroll attractor before making a transition to any one of the adjacent attractors.  We fix $N=1$ and $\alpha=7.52$ for which three-scroll chaotic attractor occurs. 
The residence  times on the left-, middle- and right-scroll attractors are calculated.  Unlike $t_{\mathrm{FPT}}$s the $t_{\mathrm{RT}}$s on all the three scroll attractors exhibit  a band-like structure over the entire possible range of values of $t_{\mathrm{RT}}$.  The middle-scroll attractor has more bands than the other two scroll attractors.  This is because a trajectory  can enter into the middle-scroll attractor  either through the bottom of the attractor (when it enters the attractor from the right-scroll attractor) or through the top of the attractor (when it  enters the attractor from the left-scroll attractor).  But a  trajectory can enter into the left(right)-scroll attractor from the middle-scroll attractor only through the bottom(top) side of the attractor.  The probability distribution curves shown in Fig.~\ref{fig9} indicates that the bands of RT are very thin.
  
\begin{figure}[!h]
\begin{center}
\epsfig{figure=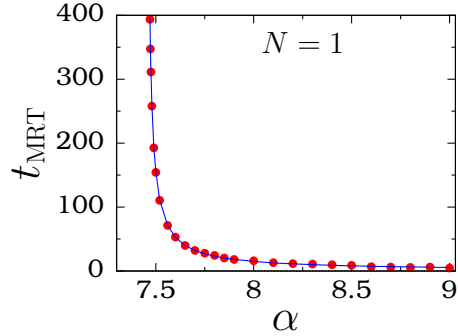, width=0.38\columnwidth}
\end{center}
\caption{Variation of  $t_{\mathrm{MRT}}$ of the middle-scroll attractor with $\alpha$ for $N=1$. $t_{\mathrm{MRT}}$ of the three  scroll attractors are all almost same.  The painted circles are numerical result while the continuous line is the best power-law fit.    }
 \label{fig10}
\end{figure}

When the value of $\alpha$ increases the band-like pattern of RT persists while the maximum value of it decreases. Figure \ref{fig10} shows mean RT (MRT),  $t_{\mathrm{MRT}}$, versus $\alpha$.  Though the number of bands of $t_{\mathrm{RT}}$ on the middle-scroll attractor is larger than the other scroll attractors the MRTs on these three attractors are almost same.  For example, for $\alpha=7.47$, $t_{\mathrm{MRT}}$ for the middle and the other two attractors are $393.73$ and $389.71$ respectively.  For $\alpha=7.52$ the values of $t_{\mathrm{MRT}}$ for these attractors are $110.54$ and $109.75$ respectively.  The difference is less than $2\%$.  Power-law dependence  of MRT on $\alpha$ with the exponent value $\approx 1$ is observed:  $t_{\mathrm{MRT}}=8.413 (\alpha - \alpha_{\mathrm{c}})^{-0.962}$.  RTs and MRT on a scroll attractor depend on $\alpha$ but independent of $N$.  Thus the MRT on a scroll attractor can be monitored by the control parameter $\alpha$.  

We believe that the study of RT and MRT on multi-scroll attractors may find practical applications.  It is noteworthy to mention that RT based detection strategies for nonlinear sensors have been proposed [Gammaitoni \& Bulsara, 2002; Bulsara {\emph{et al}}., 2003; Dari {\emph{et al}}., 2010].  This we have explored in the system (\ref{eq1}).  Often spectral  technique is used to detect the presence of weak dc or low-frequency signal.  An alternate approach in the case of low-frequency signal is the use of tuning of an internal or external noise to induce the stochastic resonance. At an optimal noise intensity in a bistable  system the signal-to-noise ratio measured at the low-frequency of the signal becomes maximum.  To detect a weak dc signal  one can use residence time asymmetry in noisy bistable devices.  The effect of an additional dc signal is to skew the potential.  In the absence of dc signal the RT distributions and MRTs in the two wells of the bistable double-well systems are identical.  They will be different in the presence of additional dc signal and the difference in the MRT is shown to be proportional to the weak dc signal [Gammaitoni \& Bulsara, 2002].  This characteristic feature can be used to identify the presence of a weak dc signal.  In the multi-scroll circuit with $f(x)$ given by Eqs.~(\ref{eq1}d-e) the MRTs of a trajectory in all the scrolls are almost identical.  This property can also be explored for weak dc signal detection as shown below. 

\begin{figure}[t]
\begin{center}
\epsfig{figure=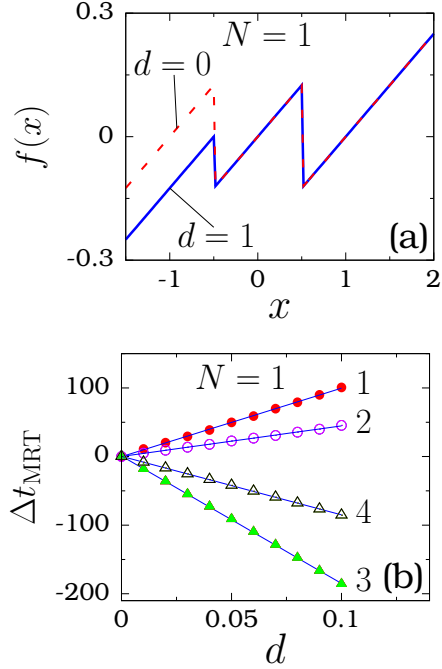, width=0.37\columnwidth}
\end{center}
\caption{(a) The graph of $f(x)$ given by Eq.~(\ref{eq6}) with $N=1$, $\xi=0.25$, $A=0.5$ and for $d=0$ and $1$.  (b) Dependence of $\Delta t_{\mathrm{MRT}}=t_{\mathrm{MRT}}(d=0) - t_{\mathrm{MRT}}(d)$ of  left - and middle-scroll chaotic attractors on $d$ for two values of $\alpha$.  Curves 1 and 2 are for middle-scroll attractor with $\alpha=7.46$ and $7.47$ respectively.  Curves 3 and 4 are for left-scroll attractor with $\alpha=7.46$ and $7.47$ respectively. The symbols are numerical data and continuous lines are the best straight-line fit. }
 \label{fig11}
\end{figure}

We introduce asymmetry in $f(x)$ given by Eqs.~(\ref{eq1}d-e) by redefining it as
\begin{subequations}
 \label{eq6}
\begin{eqnarray}
   f(x) & = & \xi x  - \xi A \sum_{j=0}^{N-1} {\bigg{\{}}
                {\mathrm{sgn}} {\big{[}}x+ (2 j+1) A {\big{]}}
                 +  {\mathrm{sgn}} {\big{[}}x- (2 j+1) A {\big{]}} + g
                     {\bigg{\}}} \;,
\end{eqnarray}
where 
\begin{eqnarray}
  g  & = & \left\{ \begin{array}{rl}
             d, &  \; \mathrm{if} \; x+(2j+1)A<0 \\ 
             0, &  \; {\mathrm{otherwise}} . \end{array}  \right.
\end{eqnarray}
\end{subequations}
Figure \ref{fig11}(a) shows the plot of $f(x)$ versus $x$ for $N=1$ with $d=0$ and $1$.  The effect of $d$ can be clearly seen.  Such forms of $f(x)$ are studied both theoretically  and experimentally [Aziz-Alaoui, 1999; Lu \& Chen, 2006; Suykens \& Vandewalle, 1993].  For $d=0$ the heights of the barriers at the breakpoints $x^*=0.5$ and $-0.5$ are same.  For $d=1$ the barrier height at $x^*=0.5$  remains same while at $x^*=-0.5$ is decreased.  This asymmetry in the barrier heights can be used to detect the weak signal by measuring the difference in the MRTs on the left- and middle-scrolls with and without the dc signal.  The MRT of a trajectory in the right-scroll remains same for $d=0$ and $d \ne 0$.  We define $\Delta t_{\mathrm{MRT}} = t_{\mathrm{MRT}} (d=0) - t_{\mathrm{MRT}} (d)$.  $\Delta t_{\mathrm{MRT}}$ on left- and middle-scrolls are calculated numerically for $\alpha=7.46$ and $7.47$ for a range of values of $d$.  Figure \ref{fig11}(b) shows the variation of $\Delta t_{\mathrm{MRT}}$ with $d$.  Interestingly, it is found to show linear variation with $d$.  From the predetermined relation between $\Delta t_{\mathrm{MRT}}$ and $d$  we can make an estimate of the value of $d$ by calculating $\Delta t_{\mathrm{MRT}}$.

\begin{figure}[t]
\begin{center}
\epsfig{figure=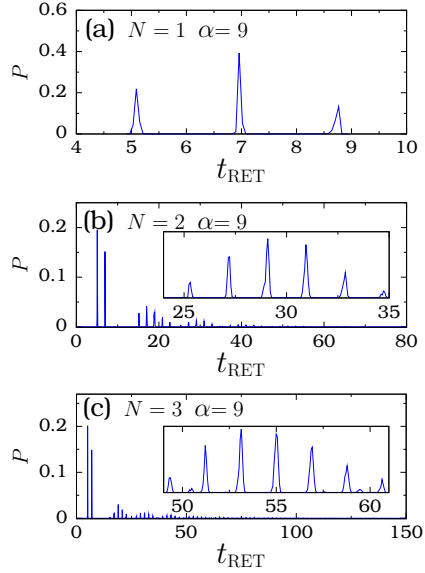, width=0.35\columnwidth}
\end{center}
\caption{Probability distribution of returned times to the middle attractor for three values of $N$ with $\alpha=9$.  }
 \label{fig12}
\end{figure}

We now turn to the discussion on returned time  to the middle-scroll attractor ($t_{\mathrm{RET}})$, the time taken by a trajectory to re-enter into the middle attractor after leaving it.  For a trajectory started with an initial condition near the origin $10^5$ RETs and the mean value of RETs denoted as $t_{\mathrm{MRET}}$ are numerically calculated for various values of $\alpha$ and $N$. For  $N=1$ three discrete bands of RETs occur.  The number of bands increases with increase in the value of $N$.  $P(t_{\mathrm{RET}})$  obtained with $10^5$ RETs for $N=1$, $2$ and $3$ are reported in Fig.~\ref{fig12}.  Distribution of RETs is different from those of FPT and RT.  The width of all the bands of RETs are very small.  In the Figs.~\ref{fig12}(b) and \ref{fig12}(c) we can clearly notice a series of Gaussian-like profile of $P$.  The first profile is predominant over the others.  The height of the profiles decays rapidly with $t_{\mathrm{RET}}$.  The above characteristics of RETs are found for higher values of $N$ also.

\begin{figure}[!ht]
\begin{center}
\epsfig{figure=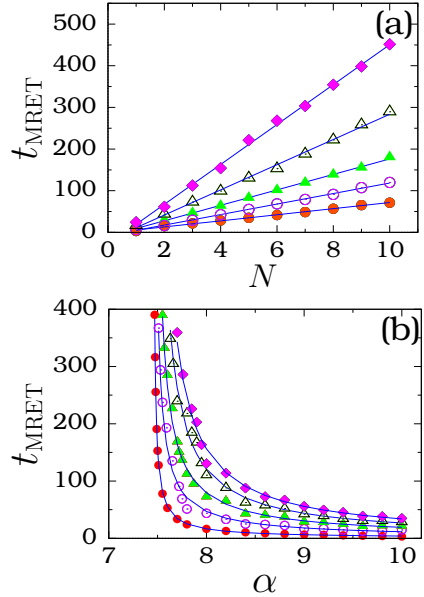, width=0.35\columnwidth}
\end{center}
\caption{Variation of mean returned time to the middle attractor with (a) $N$ and (b) $\alpha$.  In (a) the values of $\alpha$ for the top to bottom curves are $7.8$, $8$, $8.5$, $9$ and $10$ respectively.  In (b) the values of $N$ for the bottom to top curves are $1$, $2$, $3$, $4$ and $5$ respectively.  The symbols are numerical data while the continuous lines are the best fits.  }
 \label{fig13}
\end{figure}

Next, we observe different kinds of dependence of $t_{\mathrm{MRET}}$ on $N$ and $\alpha$.  In Figs.~\ref{fig13}(a) we plotted $t_{\mathrm{MRET}}$ versus $N$ for various fixed values of $\alpha$.  For each fixed value of $\alpha$ the $t_{\mathrm{MRET}}$ increases linearly with $N$.  This is because the number of scroll attractors increases linearly with $N$.  We obtained the following linear fits:

$ \alpha=7.8 \;\;:  \; t_{\mathrm{MRET}} = 47.920 N - 28.602$.

$ \alpha=8.0 \;\;:  \; t_{\mathrm{MRET}} = 30.243 N - 18.995$.
 
$ \alpha=8.5 \;\;:  \; t_{\mathrm{MRET}} = 18.557 N - 9.381$.

$ \alpha=9.0 \;\;:  \; t_{\mathrm{MRET}} = 12.639 N - 7.926$.

$ \alpha=10.0\, :  \; t_{\mathrm{MRET}} = 7.329 N - 1.668$.

\noindent The rate of divergence of $t_{\mathrm{MRET}}$ with $N$ decreases with increase in $\alpha$.  Regarding the dependence of $t_{\mathrm{MRET}}$ on $\alpha$ we notice power-law variation in Fig.~\ref{fig13}(b).   The following result is obtained:

$ N=1 :  \; t_{\mathrm{MRET}} 
         = 9.499 (\alpha - \alpha_{\mathrm{c}})^{-0.923}$.

$ N=2 :  \; t_{\mathrm{MRET}} 
         = 27.055 (\alpha - \alpha_{\mathrm{c}})^{-0.913}$.

$ N=3 :  \; t_{\mathrm{MRET}} 
         = 45.618 (\alpha - \alpha_{\mathrm{c}})^{-0.922}$.

$ N=4 :  \; t_{\mathrm{MRET}} 
         = 67.093 (\alpha - \alpha_{\mathrm{c}})^{-0.980}$.

$ N=5 :  \; t_{\mathrm{MRET}} 
         = 85.992 (\alpha - \alpha_{\mathrm{c}})^{-0.993}$.

\section{Normal Diffusion}
\begin{figure}[t]
\begin{center}
\epsfig{figure=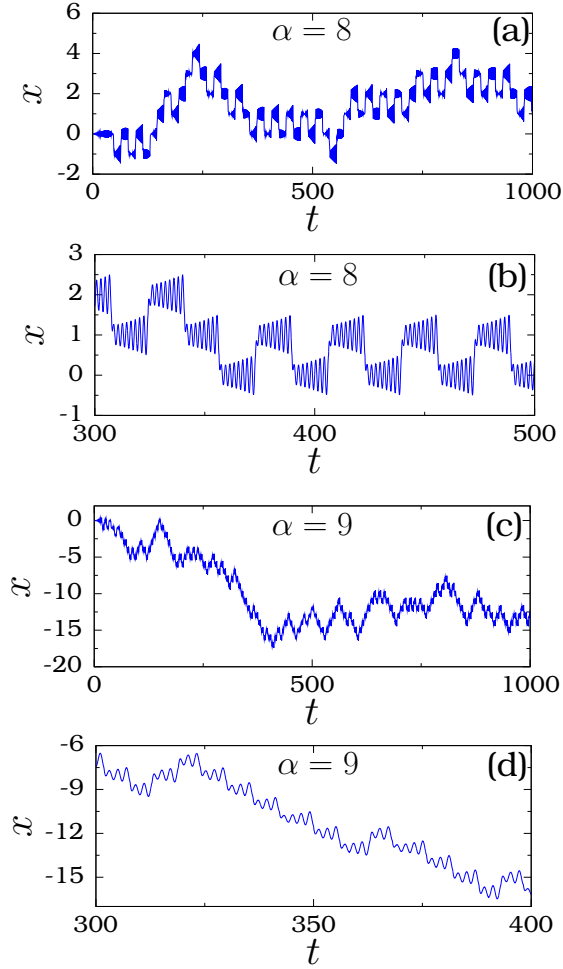, width=0.48\columnwidth}
\end{center}
\caption{$x(t)$ versus $t$ for two values of $\alpha$ of the system (\ref{eq1}) when $N=\infty$.  The subplots (b) and (d)  are magnification of detail evolution of $x(t)$ of part of the solution shown in (a) and (c) respectively.  }
 \label{fig14}
\end{figure}

In this section we report the occurrence of normal diffusion in Eq.~(\ref{eq1}).  When $N=\infty$ the function $f(x)$ has infinite  number of saw-tooth segments  and for $\alpha > \alpha_{\mathrm{c}}$ the motion is not bounded  to a finite range of $x$ in the limit $t \to \infty$.  Figure \ref{fig14} shows $x$ versus $t$  for $\alpha=8$ and $9$.  For $\alpha=8$ the divergence of $x$ is  very slow.  We can notice relatively a higher divergence  of $x$  for $\alpha=9$ in Figs.~\ref{fig14}(c) and \ref{fig14}(d).  The motion is essentially diffusive.

\begin{figure}[t]
\begin{center}
\epsfig{figure=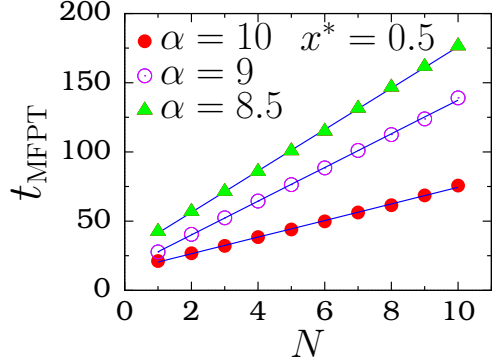, width=0.4\columnwidth}
\end{center}
\caption{Plot of  $\langle x^2(t) \rangle$ as a function of time in $\log_{10}-\log_{10}$ scale for (a) $\alpha=8$, (b) $\alpha=9$ and (c) $\alpha=10$ for the system (\ref{eq1}).  The symbols are numerical data and continuous lines are the best straight-line fits. }
 \label{fig15}
\end{figure}

In order to capture  the type of diffusion we calculate the mean square displacement over a set of $M$ initial conditions given by
\begin{equation}
 \label{eq7}
   \langle x^2(t) \rangle
     = \lim_{ \begin{array}{l} {^{M \to \infty}_{t \to \infty}} 
                \end{array} } 
        \frac{1}{M}
        \sum_{i=1}^M \left[ x^{(i)}(t) - \langle x^{(i)}(t) \rangle
           \right]^2 \;,
\end{equation}
where $x^{(i)}$ is the $i$th trajectory and $\langle x^{(i)}(t) \rangle$ is the mean value of $x^{(i)}(t)$. In our numerical simulation $M=10^4$ and the initial conditions are chosen around the origin.  The variation of $\langle x^2(t) \rangle$ against $t$ is shown in Fig.~\ref{fig15} for three values of $\alpha$.  On the $\log_{10}-\log_{10}$ scale the slopes of the best straight-line fits for $\alpha=8$, $9$ and $10$ are $\approx 1$.  In other words, $ \langle x^2(t) \rangle \sim t^{\mu}$ as $t \to \infty$ with $\mu=1$.  

A linear time dependence of $\langle x^2(t) \rangle$ is the hallmark of normal diffusion and such a diffusion process can be described by a classical random walk.  The value of the exponent $\mu$ is calculated for a range of values of $\alpha$ in the interval $[\alpha_{\mathrm{c}},11]$.  In all the cases $\mu$ is found to be $1$ and anomalous diffusion ($\mu \ne 1$) is not found.  We note that anomalous diffusion is found in several conservative systems when the phase space has accelerating modes or stochastic layers.  In these systems when a trajectory comes closer to such an accelerating mode or a stochastic layer the system is accelerated rapidly and $\langle x^2 \rangle$ diverges nonlinearly with time during the trapped times and moves chaotically otherwise.  The combination of these two types of motion leads to anomalous diffusion.   In certain dissipative chaotic diffusive systems divergence of $x$ takes place either during laminar intervals or chaotic evolution of a state variable.    In the system (\ref{eq1}) there is no laminar phase.  The growth of the state variable $x$ is due to the passage of the trajectory to the adjacent scroll attractors along one direction and also  because of the presence of infinite number of scroll attractors.  This is clearly evident in Fig.~\ref{fig14}(d). In this way diffusion in the system (\ref{eq1}) is different from the diffusion in the systems like damped and forced pendulum.

The coefficient $D_{\mathrm{c}}$  in $ \langle x^2(t) \rangle = D_{\mathrm{c}} t$ increases with increase in the value of $\alpha$.  The values of $ D_{\mathrm{c}}$ for $\alpha=8$, $9$ and $10$ are $0.018$, $0.146$ and $0.346$ respectively.  For a one-dimensional piece-wise linear periodic map Schuster and Just [Schuster \& Just, 2005] shown that the presence of diffusion implies chaotic motion and further obtained the scaling law associated with the diffusion coefficient $D_{\mathrm{c}}$.  For the conservative standard map an analytical expression for momentum distribution at time $n$ for momentum scales larger than the control parameter $K$ is obtained [Ott, 1993].  Damped oscillatory  variation of $D_{\mathrm{c}}$ with the control parameter $K$ about the quasilinear value of $D_{\mathrm{c}}$  has been reported [Rechester \& White, 1980].   

\begin{figure}[t]
\begin{center}
\epsfig{figure=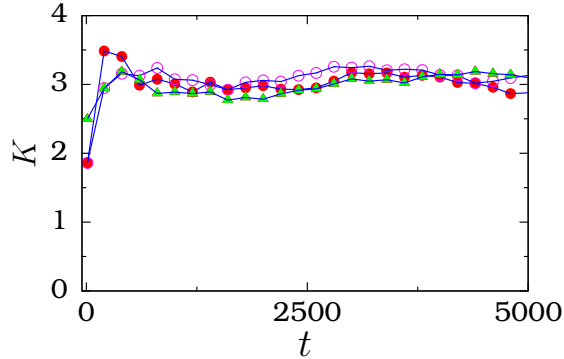, width=0.45 \columnwidth}
\end{center}
\caption{Kurtosis $K$ versus $t$ for $\alpha=8$ (marked by painted circles), $9$ (open circles) and $10$ (painted triangles) for the system (\ref{eq1}). }
 \label{fig16}
\end{figure}

Normal diffusion can be characterized by kurtosis which is defined as
\begin{equation}
 \label{eq8}
    K = \frac{ \langle (x-\langle x \rangle)^4 \rangle }
          {\langle (x-\langle x \rangle)^2 \rangle^2} \;.
\end{equation}
For normal diffusion $K=3$ in the limit $t \to \infty$.  In Fig.~\ref{fig16} the variation of $K$ is plotted for three values of $\alpha$.  For large $t$ the value of $K$ is $\approx 3$ confirming normal diffusion. 

We studied the occurrence of  diffusion in two other multi-scroll chaotic systems, one  with $ f(x ) = \sum_{j=-N}^M (-1)^{j-1} {\mathrm{tanh}} [k(x-2j)]$ [Ozoguz {\emph{et al}}., 2002; Salama {\emph{et al}}., 2003],  hyperbolic tangent  function, realizable using a multi-cycle transconductors composed of alternating D-P cells and another one proposed by Tang and his co-workers [Tang {\emph{et al}}., 2001] in which $f(x)$ is a sine function.  Normal diffusion is found in both the systems.
\section{Conclusion}
In the present work we have considered the modified Chua's circuit model equation [Eq.~(\ref{eq1})] with saw-tooth function for $f(x)$ proposed to generate multi-scroll chaotic attractors.  Attention has been paid to the study of characteristics of FPTs, RTs and RTs.   We have presented the distribution of these quantities and explored the dependence of mean values of them on the control parameter $\alpha$ and the number of scroll attractors.  From our study we can make some general  results for infinitely large number of scroll attractors.  Because the forms of the function $f(x)$ associated with all the  scroll attractors are identical the resident time characteristics of all the scroll attractors are almost the same  within the numerical accuracy.  Further, the returned times statistics of all the scroll attractors are also same.  In the system with the infinite number of scroll attractors the motion is shown to be normal diffusion type.  We wish to point out that the  multi-scroll systems and the features of multi-scroll attractors  can be useful to emulate various logic gates  and the ability to switch easily between the different operational roles.  In a recent work Murlai et al. [Murali {\emph{et al}}., 2009] have shown that the response of a bistable system with a two square waves and appropriate noise intensity produces a logic outputs (NOR/OR) and (NAND/AND).  We believe that similar results can be realized if one replaces the bistable system by the multi-scroll systems.  Further the different scrolls can be used to represent different logic states in which case MRT on a scroll attractor is important.  In the system (\ref{eq1}) MRT on a scroll can be varied over a wide range of the control  parameter $\alpha$ as shown in Fig.~\ref{fig10}.

\vskip 5pt
\noindent{\bf{Acknowledgments}} 

The authors are thankful to the referees for their suggestions which improved the quality and presentation of this paper.  One of us (SR) thanks Prof.K.P.N. Murthy for very helpful discussions. 
\vskip 5pt
\noindent{\bf{References}}
\vskip 5pt

\begin{description}
\item{}
Arena, P., Baglio, S., Fortuna, L. \& Manganaro, G.
[1996] ``Generation of $n$-double scrolls via cellular neural
networks," {\emph{Int. J. Circuit Th. Appl.}} {\bf{24}}, 241--252.
\item{}
Aziz-Alaoui, M.A. [1999] ``Differential equations with multispiral attractors," {\emph{Int. J. Bifurcation and Chaos}} {\bf{9}}, 1009--1039.
\item{}
Bilotta, E., Pantano, P., \& Stranges, F. [2007] ``A gallery of Chua attractors: Part I,"  {\emph{Int. J. Bifurcation and Chaos}} {\bf{17}}, 1--60.
\item{}
Blackburn, J.A. \& Gronbech-Jensen, N. [1996] ``Phase diffusion in a chaotic pendulum," {\emph{Phys. Rev. E}} {\bf{53}}, 3068-3072.
\item{}
Braiman, Y. \& Goldhirsch, I. [1991] ``Taming chaotic dynamics with weak periodic perturbations," {\emph{Phys. Rev. Lett.}} {\bf{66}}, 2545--2548.
\item{}
Bulsara, A.R., Seberino, C., Gammaitoni, L., Karlsson, M.F., Lundqvist, B. \& Robinson, J.W.C. [2003] ``Signal detection via residence-time asymmetry in noisy bistable devices," {\emph{Phys. Rev. E}} {\bf{67}}, 016120:1--21.
\item{}
Campos-Canton, E., Barajas-Ramirez, J.G., Solis-Perales, G. \& Femat, R. [2010] ``Multi-scroll attractors by switching systems," {\emph{Chaos}} {\bf{20}}, 013116:1--6.
\item{}
Dana, S.K., Singh, B.K., Chakraborty, S., Yadav, R.C., Kurths, J., Osipov, G.V., Roy, P.K. \& Hu, C.K. [2008] ``Multi-scroll in coupled double scroll type oscillators," {\emph{Int. J. Bifurcation and Chaos}} {\bf{18}}, 2965--2980.
\item{}
Dari, A., Bosi, L. \& Gammaitoni, L. [2010] ``Nonlinear sensors: An approach to the residence time detection strategy," {\emph{Phys. Rev. E}} {\bf{81}}, 011115:1--10.
\item{}
Gammaitoni, L. \& Bulsara, A.R. [2002] ``Noise activated nonlinear dynamic sensors," {\emph{Phys. Rev. Lett.}} {\bf{88}}, 230601:1--4.
\item{}
Harish, R., Rajasekar, S. \& Murthy, K.P.N. [2003], ``Diffusion in a periodically driven damped and undamped pendulum," {\emph{Phys. Rev. E}} {\bf{65}}, 046214:1--9.
\item{}
Kapitaniak, T. \& Chua, L.~O. [1994] ``Hyperchaotic attractors of unidirectionally coupled Chua’s circuit," {\emph{Int. J. Bifurcation and Chaos}} {\bf{4}}, 477--482.
\item{}
Khibnik, A.J., Roose, D., \& Chua, L.O. [1993] ``On periodic orbits and homoclinic bifurcations in Chua's circuit with a smooth nonlinearity," {\emph{Int. J. Bifurcation  and  Chaos}} {\bf{3}}, 363--384.  
\item{}
Lamarque, C.H., Janin, O., \& Awrejcewicz, J. [1999] ``Chua system with discontinuities,"  {\emph{Int. J. Bifurcation  and  Chaos}} {\bf{9}}, 591--616. 
\item{}
Lu, J., Chen, G., Hu, X. \& Leung, H. [2004] ``Design and analysis of multi-scroll chaotic attractors from saturated function series," {\emph{IEEE Trans. Circuits Syst.-I}} {\bf{51}}, 2476--2490.
\item{}
Lu, J. \& Chen, G. [2006] ``Generating multi-scroll chaotic attractors:  Theories, methods and applications,"  {\emph{Int. J. Bifurcation and Chaos}} {\bf{16}}, 775--858.
\item{}
Lu, J., Murali, K., Sinha, S., Leung, H. \& Aziz-Alaoui, M.A. [2008] ``Generating multi-scroll chaotic attractors by thresholding," {\emph{Phys. Lett. A}} {\bf{372}}, 3234--3239.
\item{}
Madan, R.N. ed. [1993] {\emph{Chua's Circuit: A Paradigm for Chaos}} (World Scientific, Singapore).
\item{}
Mahla, A.I. \& Badan Palhares, A.G. [1993] ``Chua's circuit with discontinuous nonlinearity," {\emph{J. Circuits, Systems and Computers}} {\bf{3}}, 231--237.
\item{}
Munoz-Pacheco, J.M. \& Tlelo-Cuautle, E. [2009] ``Automatic synthesis of 2D-$n$-scrolls chaotic systems by behavioral modeling," {\emph{J Applied Research and Technology}} {\bf{7}}, 5--14. 
\item{}
Ott, E. [1993] {\emph{Chaos in Dynamical Systems}} (Cambridge University Press, Cambridge).
\item{}
Ozoguz, S., Elwakil, A.S. \& Salama, K.N. [2002] ``$n$-scroll chaos generator using nonlinear transconductor," {\emph{Electron. Lett.}} {\bf{38}}, 685--686.
\item{}
Popescu, M.N., Braiman, Y., Family, F. \& Hentschel, H.G.E. [1998] ``Quenched disorder enhances chaotic diffusion," {\emph{Phys. Rev. E}} {\bf{58}}, R4057--4059.
\item{}
Pyragas, K. [1992] ``Continuous control of chaos by self-controlling feedback," {\emph{Phys. Lett. A}} {\bf{170}}, 421--428.
\item{}
Rechester, A.B. \& White, R.B. [1980] `` Calculation of turbulent diffusion for the Chirikov-Taylor map," {\emph{Phys. Rev. Lett.}} {\bf{44}}, 1486--1489.
\item{}
Redner, S. [2001] {\emph{A Guide to First-Passage Processes}} (Cambridge University Press, Cambridge).
\item{}
Sakthivel, G. \& Rajasekar, S. [2010] ``Diffusion dynamics and first passage time in a two-coupled pendulum system," {\emph{Chaos}} {\bf{20}}, 033120:1--9.
\item{}
Salama, K.N., Ozoguz, S. \& Elwakil, A.S. [2003] ``Generation of $n$-scroll chaos using nonlinear transconductors," {\emph{Proceedings of IEEE Symposium on Circuits and Systems}},  176--179.
\item
Schuster, H.G. \& Just, N. [2005] {\emph{Deterministic Chaos}} (Wiley-VCH, New York).
\item{}
Suykens, J.A.K. \& Vandewalle, J. [1993] ``Generation of $n$-double scrolls ($n$ = 1, 2, 3, 4, $\cdots$),"  {\emph{IEEE Trans. Circuits Syst.-I}} {\bf{40}}, 861--867.
\item{}
Suykens, J.A.K. \& Chua, L.~O. [1997] ``$n$-double scroll hypercubes in 1-D CNNs," {\emph{Int. J. Bifurcation and Chaos}} {\bf{7}}, 1873--1885.
\item{}
Tsuneda, A. [2005] ``A gallery of attractors from smooth Chua's equation," {\emph{Int. J. Bifurcation  and  Chaos}} {\bf{15}}, 1--49.
\item{}
Tang, K.S., Zhong, G.Q., Chen, G. \& Man, K.F. [2001] ``Generation of $n$-scroll attractors via sine function," {\emph{IEEE Trans. Circuits Syst-I}} {\bf{48}}, 1369--1372. 
\item{}
Tang, K.S., Man, K.F., Zhong, G.Q. \& Chen, G. [2003] ``A modified Chua's circuit with $x \vert x \vert$," {\emph{Control Theory and Applications}} {\bf{20}}, 223--227.
\item{}
Yalcin, M.E., Ozoguz, S., Suykens, J.A.K. \& Vandewalle, J. [2001] ``$n$-scroll chaos generators:  A simple circuit model," {\emph{Electron. Lett.}} {\bf{37}}, 147--148.
\item{}
Yalcin, M.E. [2007] ``Multi-scroll and hypercube attractors from a general jerk circuit using Josephson junctions," {\emph{Chaos, Solitons \& Fractals}} {\bf{34}}, 1659--1666.  
\item{}
Yu, S., Qiu, S. \& Lin, Q. [2003] ``New results of study on generating multi-scroll chaotic attractors," {\emph{Science in China}} (series F) {\bf{46}}, 104--115.
\item{}
Yu, S., Lu, J., Leung, H. \& Chen, G. [2005] ``Design and implementation of $n$-scroll chaotic attractors from a general jerk circuit," {\emph{IEEE Trans. Circuits and Systems-I}} {\bf{52}}, 1459--1476.
\item{}
Yu, S., Lu, J. \& Chen, G. [2006] ``Design and implementation of multi-directional  grid multi-torus chaotic attractors," {\emph{Proceedings of IEEE Symposium on Circuits and Systems}}, 714--717. 
\item{}
Yu, S., Tang, W.K.S. \& Chen, G. [2007a] ``Generation of $n \times m$-scroll attractors under a Chua-circuit framework," {\emph{Int. J. Bifurcation  and  Chaos}} {\bf{17}}, 3951--3964.
\item{}
Yu, S., Lu, J. \& Chen, G. [2007b] ``A family of $n$-scroll hyperchaotic attractors and their realization," {\emph{Phys. Lett. A}} {\bf{364}}, 244--251.
\item{}
Zaslavsky, G.M. [2002] ``Chaos, fractional kinetics, and anomalous transport," {\emph{Phys. Rep.}} {\bf{371}}, 461--580.
\item{}
Zhong, G.Q., Man, K.F. \& Chen, G. [2002] ``A systematic approach to generating $n$-scroll attractors," {\emph{Int. J. Bifurcation  and  Chaos}} {\bf{12}}, 2907--2915. 

\end{description}

\end{document}